\documentclass[sigconf,nonacm]{acmart}
\usepackage{graphicx}
\usepackage{subcaption}

\AtBeginDocument{%
  \providecommand\BibTeX{{%
    \normalfont B\kern-0.5em{\scshape i\kern-0.25em b}\kern-0.8em\TeX}}}

\copyrightyear{2024}
\acmYear{2024}
\setcopyright{cc}
\acmConference[MOCO '24]{9th International Conference on Movement and Computing}{May 30-June 2, 2024}{Utrecht, Netherlands}
\acmBooktitle{9th International Conference on Movement and Computing (MOCO '24), May 30-June 2, 2024, Utrecht, Netherlands}
\acmDOI{10.1145/3658852.3659075}
\acmISBN{979-8-4007-0994-4/24/05}

\acmConference[MOCO '24]{Movement Computing}{May 30-June 2, 2024}{Utrecht, Netherlands}
\begin{document}

\title{Visual instrument co-design embracing the unique movement capabilities of a dancer with physical disability}

\author{Sam Trolland}
\orcid{0000-0002-3337-1006}
\affiliation{
    \institution{\textit{SensiLab, Monash University}}
    \city{Melbourne}
    \country{Australia}
}

\author{Melinda Smith}
\orcid{0009-0007-9030-3708}
\affiliation{
    \institution{\textit{SensiLab, Monash University}}
    \city{Melbourne}
    \country{Australia}
}

\author{Alon Ilsar}
\orcid{0000-0002-1225-2497}
\affiliation{
    \institution{\textit{SensiLab, Monash University}}
    \city{Melbourne}
    \country{Australia}
}

\author{Jon McCormack}
\orcid{0000-0001-6328-5064}
\affiliation{
    \institution{\textit{SensiLab, Monash University}}
    \city{Melbourne}
    \country{Australia}
}

\renewcommand{\shortauthors}{Trolland, et al.}

\begin{abstract}
This paper explores the design of an expressive visual instrument that embraces the unique movement style of a dancer living with physical disability. 
Through a collaboration between the dancer and an interaction designer/visual artist, the creative qualities of wearable devices for motion tracking are investigated, with emphasis on integrating the dancer's specific movement capabilities with their creative goals.
The affordances of this technology for imagining new forms of creative expression play a critical role in the design process. 
These themes are drawn together through an experiential performance which augments an improvised dance with an ephemeral real-time visualisation of the performer's movements.
Through practice-based research, the design, development and presentation of this performance work is examined as a `testbed' for new ideas, allowing for the exploration of HCI concepts within a creative context.  
This paper outlines the creative process behind the development of the work, the insights derived from the practice-based research enquiry, and the role of movement technology in encouraging new ways of moving through creative expression.
\end{abstract}



\begin{teaserfigure}
  \includegraphics[width=\textwidth]{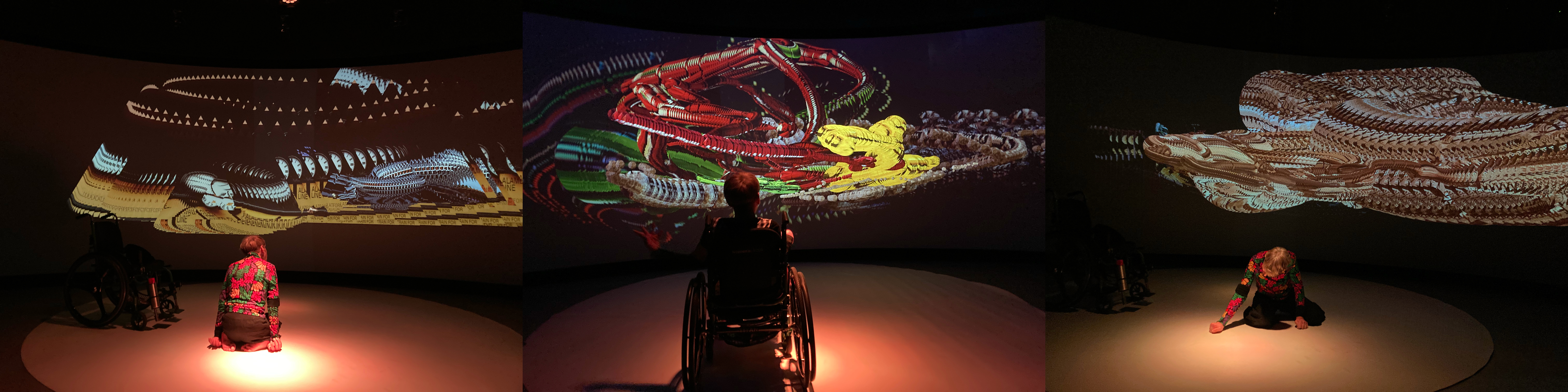}
  \caption{Interactive visualisation system responding to a performer's movements during a rehearsal session.}
  \Description{Three images of a dancer in a seated pose with an abstract visualisation projected on the wall behind them. In each image the performer is in a different position on the floor seated nearby their wheelchair.}
  \label{fig:teaser}
\end{teaserfigure}

\maketitle

\section{Introduction}
Universal design, often used interchangeably with design-for-all, is defined as the design of systems that can be used, to the greatest extent possible, by all people, without having to be adapted to specific individuals or groups \cite{catanese2012thomas, Persson2015}.
When designing a new interactive digital instrument for creative expression, there is a tendency to `universally' design the interface around a style of interaction that will engage the largest number of possible users.
This approach ensures the best chance of wide adoption but comes at the cost of excluding potential users that are not able to interact in the way the interface has been designed.
When it comes to movement computing, this divide is particularly experienced by people living with physical disability who can face barriers that exclude them from using traditional interfaces for creative expression.
Acknowledging that everyone’s range and style of movement is different, in this research
we use participatory design methods to meet the specific needs and expectations of a dancer living with physical disability; applying their knowledge and experiences throughout the design process and designing around their specific movement capabilities \cite{sanders2002user}.
In doing so, we aim to develop a truly expressive interaction system that responds to the unique qualities of an individual’s style of movement, creating new pathways for the performer to express themselves through technology.

We approach these research aims through the creation of a contemporary performance piece that augments improvised dance with an ephemeral visualisation of the performer's movements. 
Adopting a practice-based research methodology \cite{Candy06}, the creation of this performance work is used as a ‘testbed’ for new ideas, exploring Human-Computer Interaction (HCI) concepts within a creative context. 
This work details the creation of a bespoke visual instrument by exploring the creative qualities of motion-tracking wearable devices towards enhancing the performer's existing dance practice with new forms of creative expression. 

By designing movement-based interactions around a particular individual’s range and style of movement, we aim to afford that individual the widest range of creative expression the interaction technology can offer. 
This approach is explored as a method to create an inclusive interaction experience for the performer, allowing them to engage with the work through their natural style of movement rather than requiring the performer to adapt their style of movement to the capabilities of the technology.
Rigid systems that require a performer to adapt their movement style to a fixed interaction language can pose barriers for persons living with physical disability, who may not be comfortable or physically able to interact in the way the system has been designed. 
This highlights the importance of designing for the performer's specific movement capabilities and needs. 
The degree of personalisation that is possible with movement interaction systems therefore allows the interaction experience to be crafted for an individual’s specific range and style of movement. 

This paper outlines the collaborative creative process employed in the development of the performance piece; describing how the analysis of movement data was used in an iterative design cycle towards the creation of inclusive and expressive movement visualisations.
Insights drawn from observation and performer feedback are outlined and discussed, demonstrating the role of movement technology in fostering new ways of moving both within this performance piece and within the performer's daily life. 
The presence and positioning of wearable devices was found to encourage the performer to explore expressive movements with parts of their body over which they have limited control, or that are often overlooked in their other dance performances. 
The insights gained from this creative exploration of HCI concepts can be generalised for application by other creative practitioners aiming to develop expressive movement interaction experiences.

\section{Related Work}
In the design of gestural systems for HCI, there is a tendency for designers to create systems around their own preferences \cite{RienerA.2012GIiV}.
This approach has been described as \textit{``counter-intuitive''} \cite{10.1007/978-3-030-62466-8_8}, as it limits the interaction to a set of \textit{``discrete gestural commands''} \cite{Wexelblat95} that can be recognised by the interaction technology.
It involves defining a restrictive set of gestures, requiring the performer to learn how to perform these gestures in a way that is recognisable by the system.
This method approaches the design of the interaction experience giving primacy to the technology; assuming the performer can adapt their style of movement in order to perform the gesture in a way the system can recognise. 
To counter this method, we approach the interaction design by focusing on the experience of the human within the system; designing the interface to allow a performer to move and gesture in a manner that is most natural for them.
This participatory design approach, which involves including the end users within the development of a interface, is first described in \cite{schneiderman1992designing}, and is later described \textit{``critical''} for the design of gestural interfaces \cite{RienerA.2012GIiV}.

Interfaces designed to use movement of the body as a mode of interacting with digital systems rely on either external detection systems (vision or depth sensing technology) or a wearable device that can sense and quantify movement \cite{Honauer18}. 
Vision based systems have been described as a non-intrusive gestural interaction technology \cite{Wachs11}, capable of pose detection \cite{Moeslund06} but suffering from limitations in accuracy, processing speed and generality \cite{Erol07}.
Recent advances in AI have greatly improved upon these limitations in vision-based systems, however they still rely on certain lighting conditions and direct line-of-sight to operate which can prove a major limitation when designing a system for use in a live performance setting.
Wearable technologies can overcome a number of these limitations by situating a wireless device on the performer's body that measures and transmits movement information.
The Inertial Measurement Unit (IMU) is a common technology used within wearable gestural interfaces to measure movement characteristics \cite{Benbasat01}. 
The wearable device used in this research, described in \cite{Trolland2022AirSticks}, utilises an IMU sensor capable of detecting orientation and linear acceleration. 
This wearable device uses Bluetooth to wirelessly transmit this information to a nearby computer for further analysis and visualisation.
Originally designed as a gestural musical instrument, in this work we extend prior research by exploring the creative possibilities of this device as an interface for visual expression in a contemporary dance performance.

The creative qualities of IMU-based gestural devices have been explored within the movement computing community for both sonic and visual expression \cite{Schlegel17,Hoelzl19,Honauer18}. 
Hoelzl et al. \cite{Hoelzl19} detailed the creation of a `non-trivial', wireless, handheld musical instrument that was iteratively developed through a series of performance experiments and user tests.
The instrument's ability to create complex sounds with simple movements allows players to focus on listening and intuitive motion.
In Schlegel et al. \cite{Schlegel17}, `ordinary objects' were augmented with IMU sensors to explore the expressive potential of these objects across various artistic disciplines including dance, music and visual art.
In Honauer et al. \cite{Honauer18}, four wearable IMU sensors are integrated into a performer's dance costume with a flexible interface connecting the performer to the stage environment, allowing their movement to affect changes in interactive media such as stage lighting and sound. 
This work extended the interface for two-way communication, allowing the performer to receive external inputs from an off-stage operator that could be experienced through vibrational motors within the costume.  
This builds on earlier work by Hallam et al. in \cite{Hallam14} where four accelerometers were embedded into a dance instructor's costume  as a training aid for adult beginners starting ballet.
In this work the costume was augmented with LED lights, providing visual feedback as a communication tool to assist the beginners to follow the instructor's movements.

Most relevant to our research, motion capture technology has been used to detect, anaylse and characterise gestures of dancers with physical disability, particularly in the field of manual wheelchair dance \cite{Luna22}.
Accessible motion-capture video games for wheelchair users have also been developed, some for leisure \cite{gerling2015designing, de2012jewheels} and others for rehabilitation \cite{alarcon2020upper}.
Visualisations of the movements of dancers with physical disability have also been created, though the visualisations were used to provide \textit{``expressive information for movement exploration and quality assessment''} \cite{Xie23}, rather than within a performance context.

It this worth noting that the artist collaborating on the performance described in this paper uses a manual wheelchair in their dance practice; however, they also incorporate floor work, dancing in, around and with their manual wheelchair in different parts of the work.
Prior work conducted with this artist has focused on developing their music practice, utilising a gestural musical instrument to convert their movement into sound. 
Designers in the field of Accessible Digital Musical Instruments (ADMI) occasionally utilise motion capture technology for visualisations in the development of accessible musical instruments; however the visualisations are often used as feedback for the performer as opposed to being an expressive device presented to the audience during a live performance \cite{Frid19}.

\section{Creative Process}
The creative process in this project was structured around the goal of co-developing a performance piece that augments an improvised dance with an interactive visualisation of movement. 
From an interaction design perspective, the aim was to create a visualisation that responds to the style of movement that is natural within the performer's practice. 
To achieve this the interaction system was iteratively designed and refined through a series of joint rehearsal sessions.
By including the performer in the iterative design process, we were able to create an interaction system that was tailored to the unique qualities of the performer's movement capabilities, rather than requiring the performer to adapt their style of movement to accommodate the technology. 

\subsection{Description of Performance Piece}
\begin{figure*}[t!]
    \centering
    \begin{subfigure}[t]{0.5\textwidth}
        \centering
        \includegraphics[width=\textwidth]{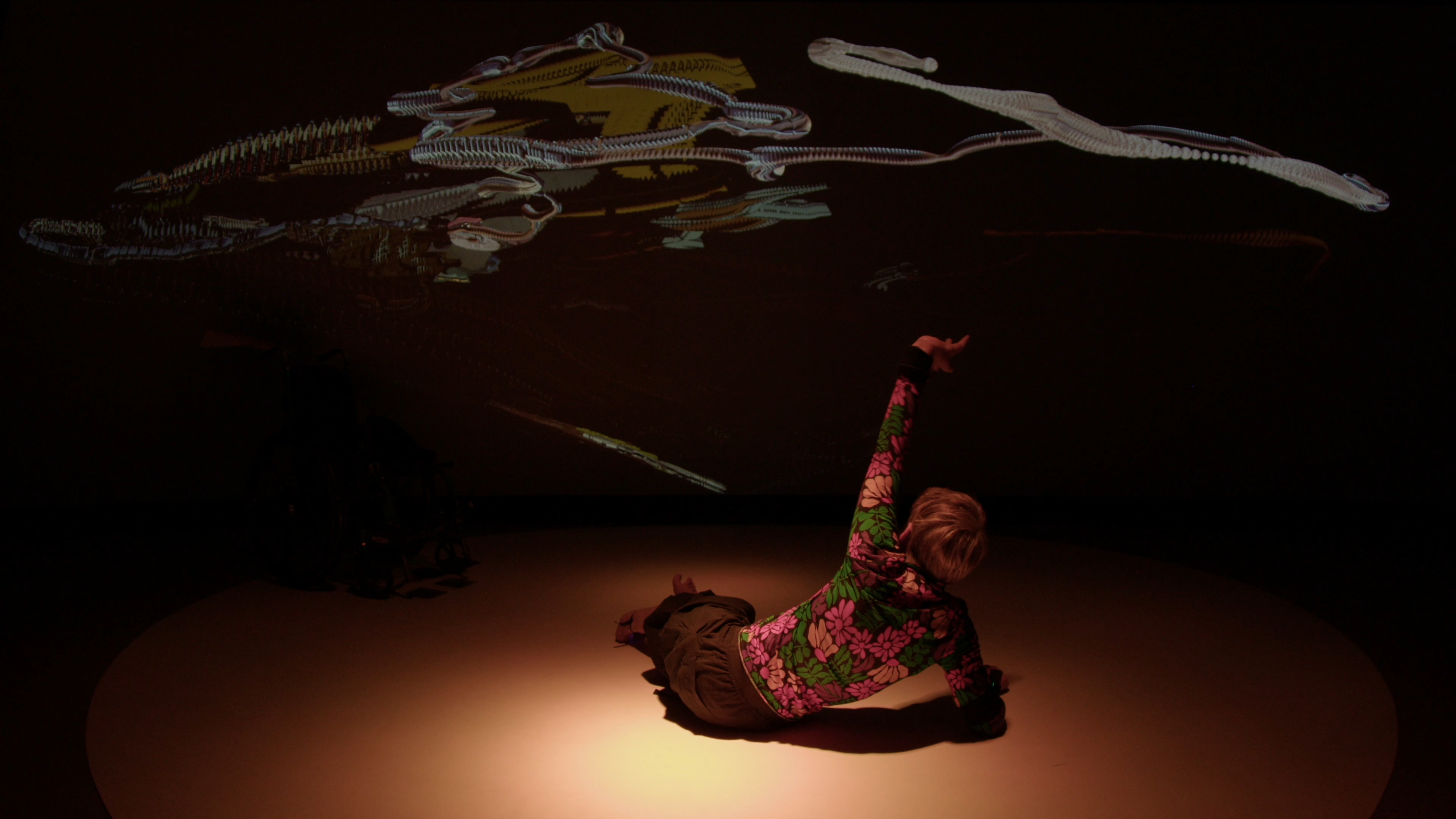}
        \caption{Visualisation resulting from slow, smooth, movements.}
    \end{subfigure}%
    ~ 
    \begin{subfigure}[t]{0.5\textwidth}
        \centering
        \includegraphics[width=\textwidth]{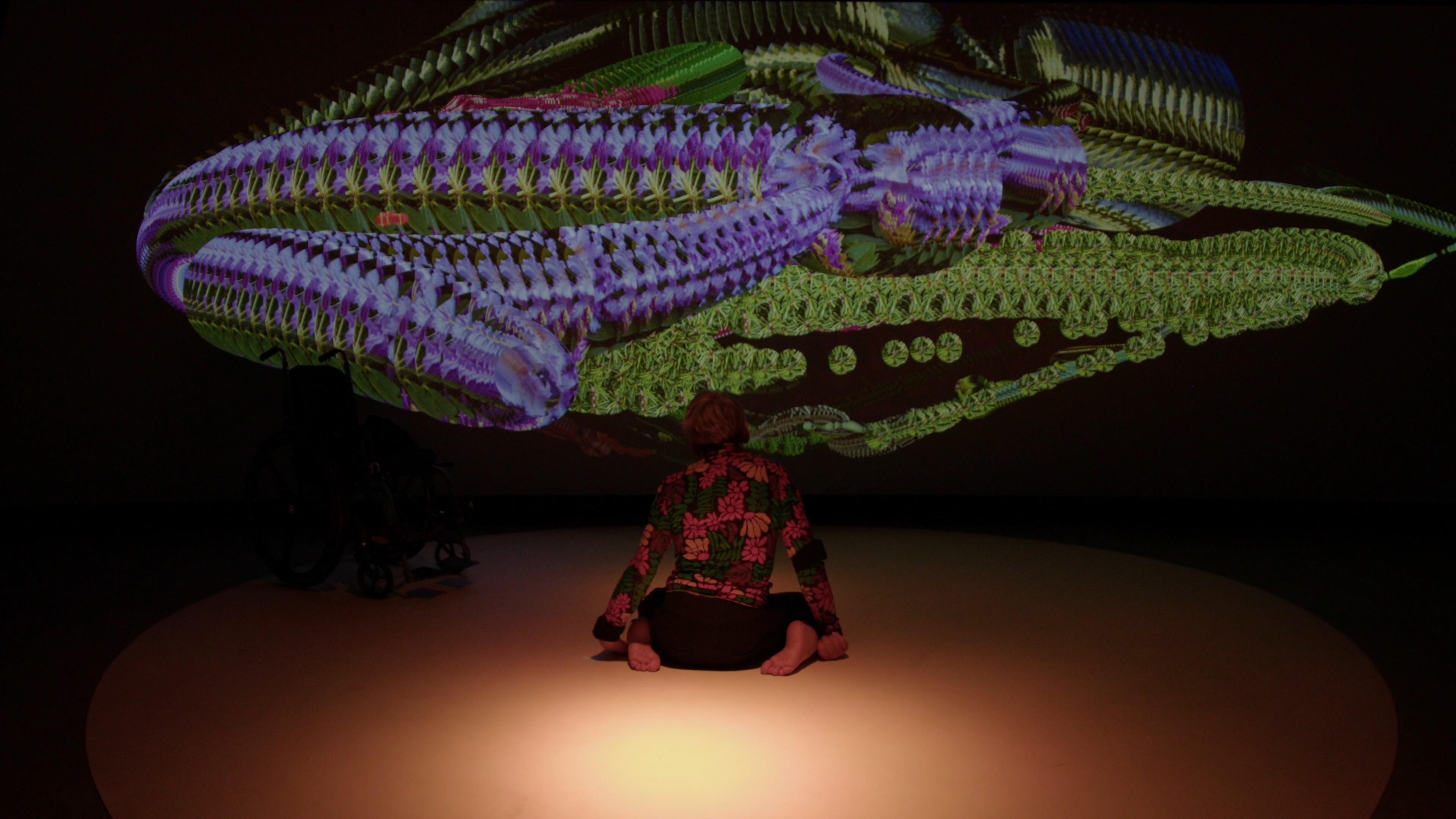}
        \caption{Visualisation resulting from fast, staccato, movements.}
    \end{subfigure}
    \caption{Performer's interactions with the visualisation during the performance showing.}
    \Description{Two images of a dancer in a seated pose with an abstract visualisation projected on the wall behind them. In the first image the visualisation is made up of thin lines and in the second image the visualisation is made up of thicker lines with brighter colours.}
    \label{fig:SlowFastInteractions}
\end{figure*}

The performance piece was presented in Melbourne on the 28th of September 2023, and comprised a 30-minute improvised dance solo augmented with an ephemeral visualisation of the performer's movements. 
Video documentation of the full performance is available to download \cite{Trolland2024} or stream\footnote{\url{https://youtu.be/m09HIKOAepk}}.
The performance space, shown in Figure \ref{fig:teaser}, was enclosed by a 3m circle of vinyl flooring, a surround sound system and a projection screen backdrop upon which the visualisations were presented. 
The piece delved into themes centred on urban and natural environments, examining the effects these distinct surroundings have on the performer, and exploring the emotional responses together with the resulting movements evoked by each setting.
These themes were explored through dance, sound, and interactive visual elements that provide a sensory augmentation of movement. 

The interactive visualisation was designed to give an experience of `painting with light', with the performer fluctuating between moments of movement and creation followed by moments of stillness and reflection.
The visualisation system was built upon movement-based manipulation of photographic textures that represented the themes of the piece. 
Guided by the performer's movements, these images navigate the digital canvas leaving behind a trail of colour and texture in an abstract visualisation of the performer's movements.
The movement of a single photo or texture across the screen is used as a metaphor for a `brush stroke' on a canvas, with the trails layering atop one another to create complex visual forms. 
An audio composition was created to complement the piece, comprising field recordings gathered from urban and natural environments.

During the performance, the performer wore four wearable devices, each controlling its own `brush' on the digital canvas.
In response to the performer's movements, these four brush strokes dynamically traverse the digital canvas in real-time, producing abstract forms that creatively visualise the performer's movements.
These visual forms slowly fade to black over time, revealing a fresh canvas for the performer to fill again. 
Slow, smooth movements create a thin trail of colour on the canvas, reminiscent of a fine single-strand brush, shown in Figure \ref{fig:SlowFastInteractions}(a). 
Quick, staccato movements produce broad brush strokes, unveiling additional textural details from the underlying photograph, as shown in Figure \ref{fig:SlowFastInteractions}(b).

The improvised performance was structured around a progression from an city environment, a busy place where the performer describes feeling \textit{``cold''} and \textit{``isolated''}, flowing into a natural environment where the performer feels more \textit{``connected rather than disconnected''}.
This was conveyed by the progression of the sonic and visual textures throughout the piece, augmenting the performer's movements to create an immersive and experiential performance for both the performer and the audience.

\subsection{Iterative Development Cycle}
The performance emerged through an iterative cycle of development, evaluation and reflection. 
The collaborative process followed in the creation of this piece, involved a series of nine rehearsal sessions, with periods of interaction and visualisation system development in-between each session.
Throughout the rehearsal sessions, we explored movement, dance and audio-visual elements that respond to the themes of the piece.

\subsubsection{Movement-led Development}
We started with a preliminary meeting to discuss our creative goals from the collaborative project, the themes we wanted to explore though the combination of movement and technology, the scope of the project including frequency and length of rehearsal sessions, and the format of the performance. 
During this session we discussed the creative potential of the wearable devices; how many we wanted to use and the possible locations they could be worn on the performer’s body. 
As the wearable devices respond to orientation and acceleration, placing them on the extremities such as ankles and wrists creates the most dynamic interactions.
In previous interactive works we have let this consideration guide the choice of sensor placement on the body. 
However, in this collaboration the performer was intrigued by exploring alternative possibilities and how the placement of the sensors influenced their movement. 
We decided to work with four wearable devices, one on the right upper arm, one on each ankle, and one on the left wrist.
During the first rehearsal we started with movement only: no visualisations or interactive technology. 
This gave the performer an opportunity to move naturally, unencumbered by technology and the influence of the visualisation on their movement decisions. 

\subsubsection{Interaction and Visualisation System Development}
The creation of the interaction and visualisation systems primarily occurred in-between the rehearsal sessions, and was led by data captured during the previous rehearsals.
Analysis of the video and gesture recordings, further described in Section \ref{analysis}, revealed insights into the aspects of the performer's movements that could be sensed and measured by the wearable devices. 
The learnings from this analysis were applied in the creation of prototype interaction and visualisation systems tailored to the performer's distinctive style of movement, in line with the creative imperatives of the piece.

\subsubsection{Rehearsal Sessions: Evaluation and Exploration}
The rehearsal sessions provided an opportunity to explore and evaluate the prototype interaction and visualisation systems in practice. 
These sessions would involve periods of movement exploration, followed by reflection and discussion; providing opportunities to fine tune interaction parameters and conceive of new visualisation concepts to explore in future development cycles. 

\subsection{Analysis of Movement Data} \label{analysis}

Movement data captured during the rehearsal sessions was analysed to observe how the performer's movements were expressed in the motion data captured by the wearable sensors. 
Video and gestural data recordings captured at the same moment were compared side-by-side to observe movement qualities in the video recordings and their associated representation within the gestural data. 
This process quantified the performer's movement characteristics expressed at each wearable device placement, and allowed for fine-tuning of the interaction parameters to ensure each sensor location is capable of achieving the full range of creative expression the visualisation has to offer.

The movement data captured by each wearable device includes the device's orientation and linear acceleration.
The orientation data identifies the direction the sensor is facing in 3D space. 
The linear acceleration data shows the instantaneous acceleration of the device along the three axes local to the sensor's frame of reference. 
When worn on the performer's body, this information can be used to quantify the direction and extent of movement at that particular part of the body at any given moment.

Analysis of the linear acceleration data revealed that the amount of movement varies significantly between each of the sensor locations.
Aggregating the absolute value of the linear acceleration across all three dimensions over a short period of time is used to represent a quantity referred to as acceleration `Energy' \cite{Trolland2022AirSticks}. 
Energy can be built up over periods of consistent, fast, movements and changes in direction. 
This Energy value slowly falls back down during periods of slow, smooth, movements and moments of stillness. 
The time period used when calculating Energy affects the rate at which the value builds-up and drops-off.
This parameter was fine-tuned while the performer interacted with the visual system, with different values chosen to fit the movement characteristics expressed at each sensor location.

\begin{figure*}[h]
  \centering
  \includegraphics[width=\linewidth]{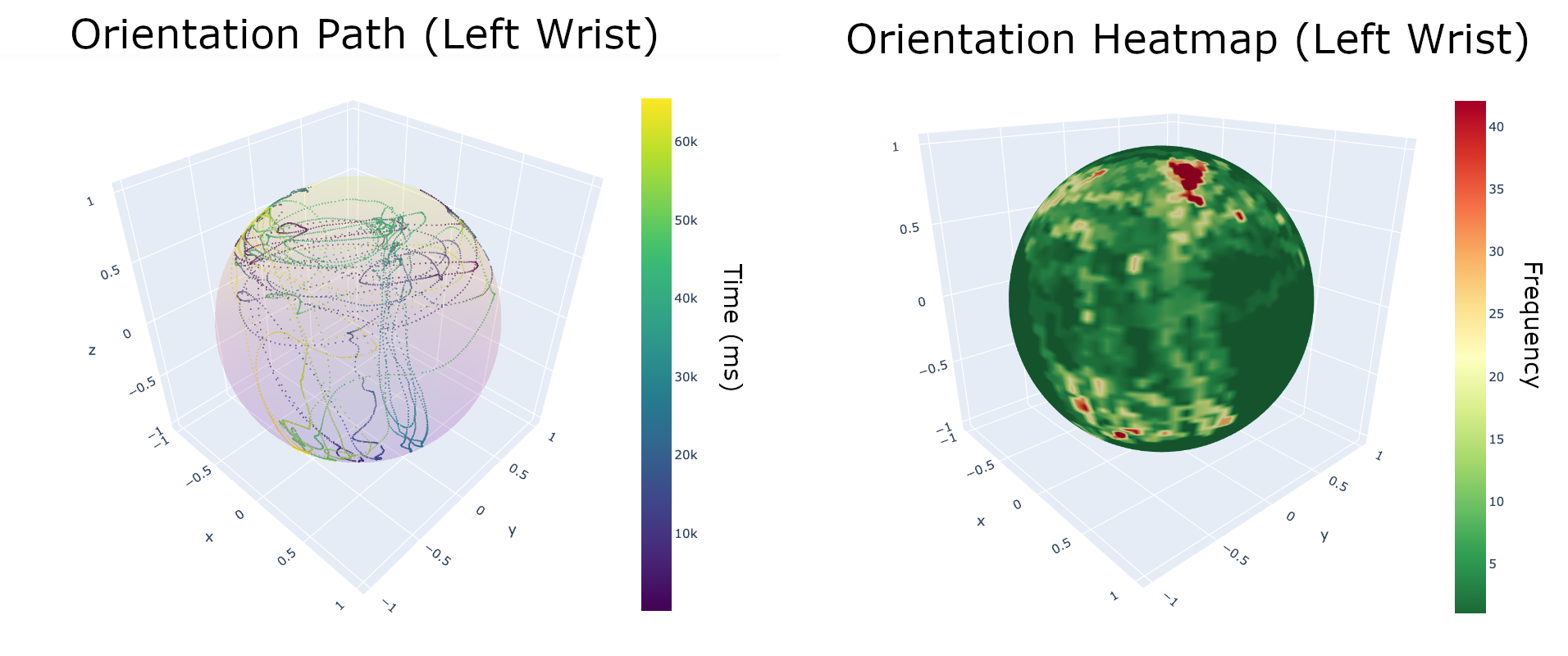}
  \Description{Two scatter plots showing a unit sphere on 3D axes. The first plot shows a thin line representing the orientation path of the performer's left wrist during a rehearsal session. The second plot shows a green sphere with yellow and red regions representing a heat map of orientations that were more common in the rehearsal recording.}
  \caption{Analysis of orientation data recorded during a rehearsal session.}
  \label{fig:orientation}
\end{figure*}

The range of the performer's movement was measured by analysing the orientation data at each sensor location. 
The path of the performer's interactions during a rehearsal session were recorded and plotted as a 3D scatter plot shown in Figure \ref{fig:orientation}. 
This plot shows the dynamic nature of the performer's movements, and was used to quantify the upper and lower bounds of the range of movement available at each sensor location. 
The path of the orientation data was further analysed to identify regions of higher and lower activity, visualised as a heat map in Figure \ref{fig:orientation}. 

\subsection{Visualisation of Performer's Movements}
The creative development of the interactive visualisation involved building a system to express aspects of the movement data in line with the artistic vision of the piece. 
The themes of the piece were realised through a dynamic display of textural photographs that captured the essence of both the urban and natural environments. 
Three hundred high-resolution, close-up photographs captured by both the performer and the interaction designer were used as the foundation of the visual aesthetic of the piece. 

Within the performance, the visualisation progressed through a series of `scenes', each comprised of around 20 photographic textures of similar visual content. 
The piece began with textural photographs of concrete floors and walls and progressed through scenes of cityscapes and buildings before moving into photographs of nature; the forest floor, bark, leaves and flowers in bloom. 
The progression through the scenes was timed to flow alongside the audio-composition created for the piece.
It was initially envisioned that the performance audio would be interactive, with elements such as volume or playback speed influenced by the performer's movements. 
However, after exploring this concept within rehearsal sessions, it was decided that a pre-recorded composition was preferred over interactive audio as this offered prompts to the performer at key moments and provided structure to the improvised dance piece.
While the timing of this progression remained the same each time the piece was performed, the textures assigned to each brush stroke were chosen at random from the current scene's collection. 
The choice to include randomness within the visual progression of the piece emerged through the rehearsal process, as it added spontaneity and enhanced the improvisational nature of the performer's practice. 

In addition to the photographic textures of urban and natural environments, the performer brought in three of their own paintings that they wished to include within the work.
These paintings were digitised and included in the visualisation to signify key moments to the performer as they performed the piece. 
At these key moments, all four of the available brush strokes were assigned the same painting texture, allowing the performer to fill the canvas with a digital reinterpretation of their artwork as shown in Figure \ref{fig:Paintings}. 

To achieve the creative goal of `painting with light' the direction of the performer's movements was used to create `brush strokes' by dragging a texture across the digital canvas.
The visualisations responded to the performer's movements in real-time and were presented on a projection screen at the rear of the stage.
A mapping process translated the performer's movements in 3D space into the 2D space of the digital canvas.
When the sensor devices were oriented towards stage left or stage right, the associated brush stroke would be positioned on the matching side of the digital canvas. 
Similarly, movements orienting the sensors towards the ceiling or the floor would be visualised on the top or bottom of the canvas respectively. 
These two axes were used as the boundaries of the orientation interaction, with symmetry built into the third axis to position the brush stroke in the centre of the canvas when the performer's movements faced the centre of the projection screen or the centre of the audience. 
This interaction was designed to allow the performer to move between moments facing the projection screen, concentrating on crafting visual forms to share with the audience, and moments facing the audience and \textit{``just moving''}, later turning back to see the creations they had made.

Dynamics in the visualisation respond to the pace of the performer's movements, utilising the Energy value at each sensor location to influence the size and detail of the associated brush stroke.
The range of Energy values at each sensor location were mapped within the visualisation to ensure the full range of brush stroke dynamics for each placement.
Moments of stillness were identified when Energy values remained below a threshold for a given period of time.
During these moments of stillness, the texture assigned to the given brush would cycle to a new photograph from within the current scene.
This interaction was designed as a metaphor for a painter dipping their brush in another pot of paint before returning to the canvas.

\section{Insights}

\subsection{Visual augmentation of movement}
\begin{figure*}[t]
    \centering
    \begin{subfigure}[t]{0.43\textwidth}
        \centering
        \includegraphics[width=\textwidth]{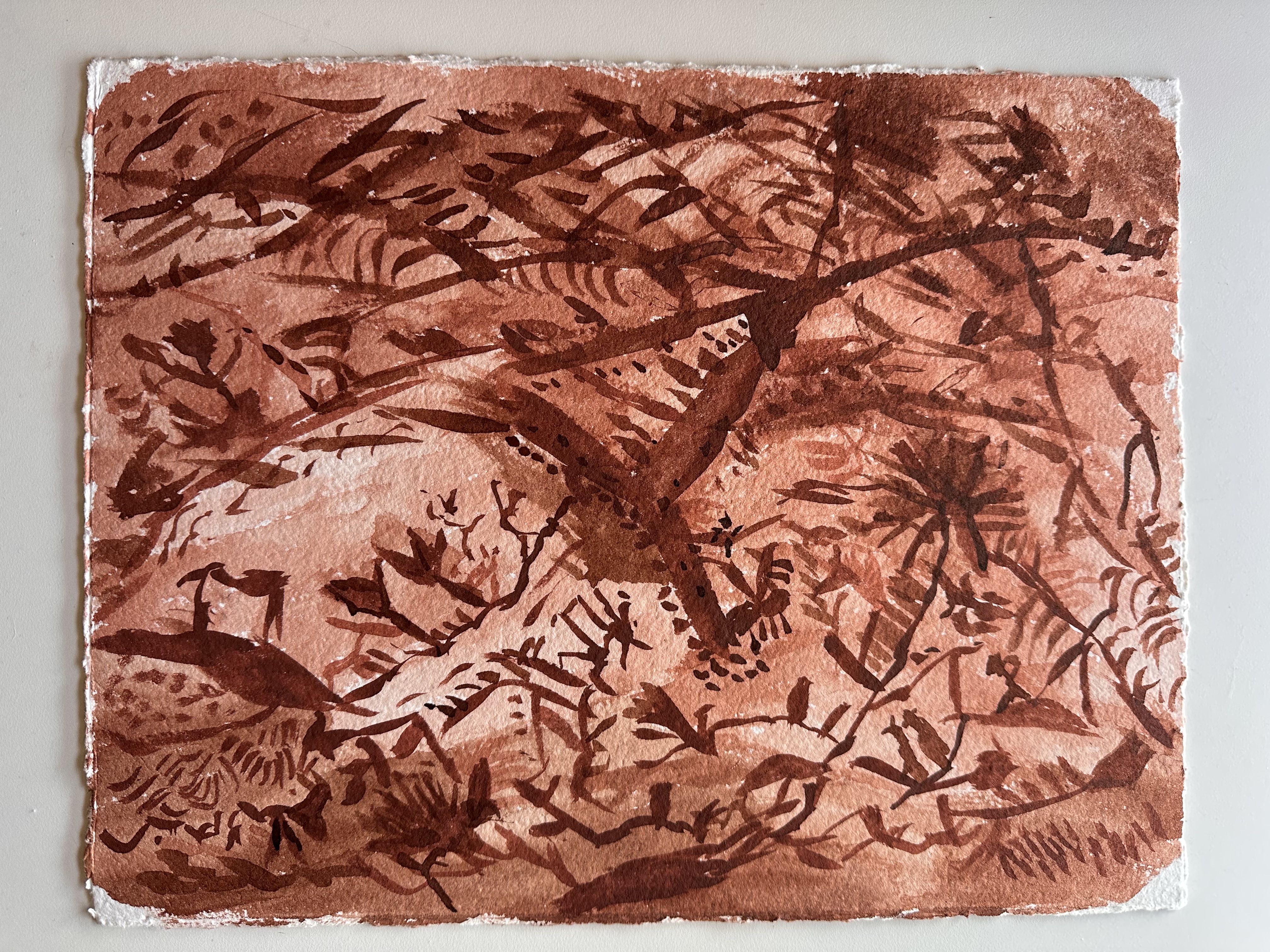}
        \caption{Performer's painting digitised for the piece.}
    \end{subfigure}%
    ~ 
    \begin{subfigure}[t]{0.57\textwidth}
        \centering
        \includegraphics[width=\textwidth]{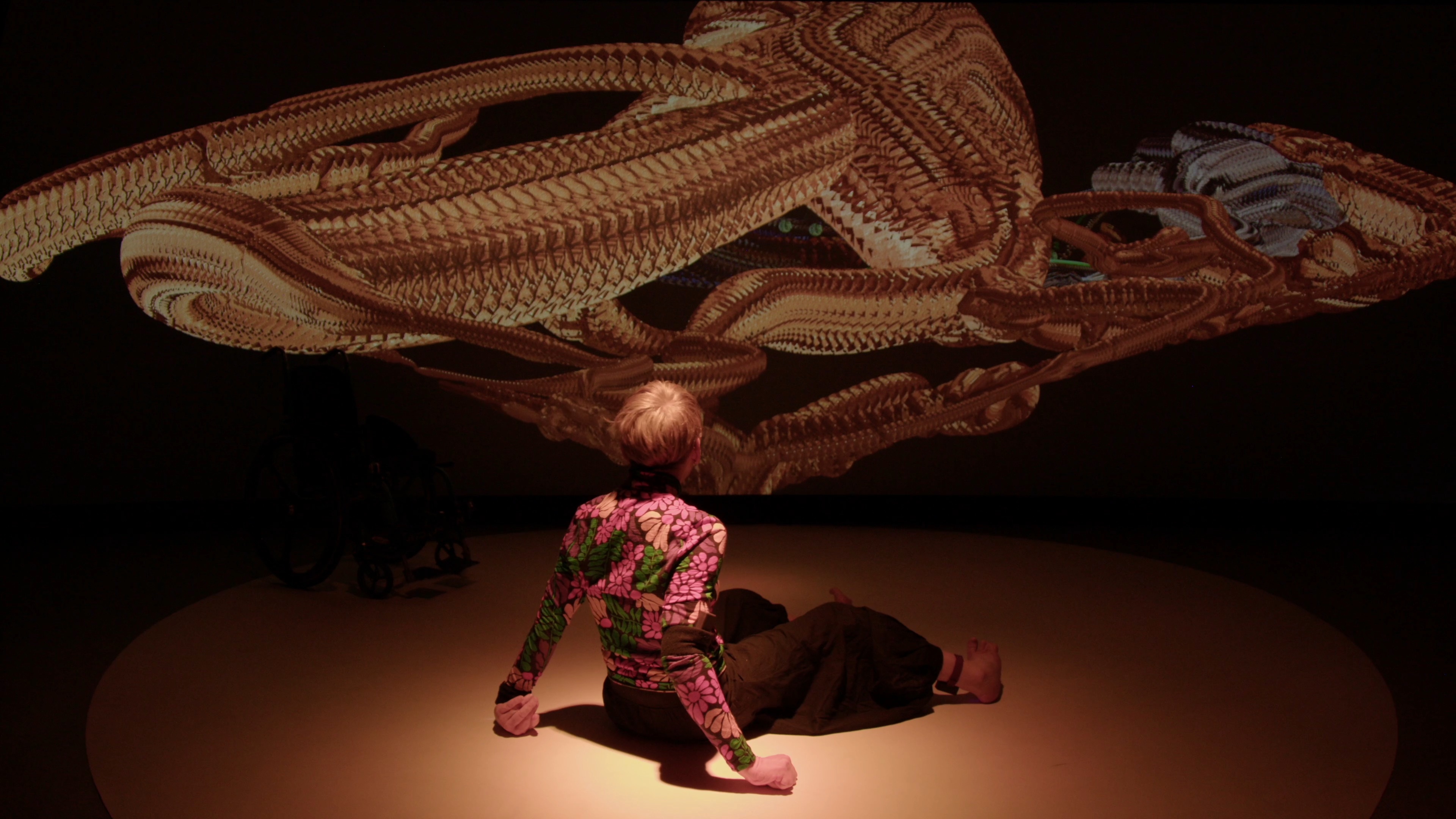}
        \caption{Reinterpretation of the painting during the performance.}
    \end{subfigure}
    \caption{Performer's interactions with digital reinterpretations of their paintings during the performance showing.}
    \Description{Two images side-by-side. The first image shows a water-colour painting with earthy tones with abstract lines and figures reminiscent of nature. The second image shows the performer in a seated pose with an abstract visualisation projected onto the wall behind them. The visualisation is made up of the same earthy tones and textural elements, reinterpreted into new shapes by the performer's movement.}
    \label{fig:Paintings}
\end{figure*}

By designing the visualisation system to respond to the performer’s unique movement style, as expressed through the gestural data recorded during the rehearsal sessions, the interactive visualisations that emerged from this collaborative process were shown to successfully enhance the performer's dance practice through sensory augmentation of movement. 
The project demonstrated that by augmenting the performer's dance practice with a movement-based interactive visualisation that was designed for, and with, the performer, an experiential performance was produced opening new avenues for the performer’s creative expression.
In an interview following the performance, the performer noted that the development and showing of the piece had allowed them to express themselves in ways they had not considered before the start of the project.
Specifically, the presence of the visualisation within the performance inspired them to think about different ways of moving, and to view their own creative practice in new ways. 
The performer stated that by the time of the first public performance \textit{``I wasn’t only a dancer, but I was an artist, a painter''}, explaining \textit{``that really opened up my imagination''}.
The performer explained that the presence of the visualisation was like \textit{``a dance partner''}, stating that it had an influence on their movement decisions during the performance and \textit{``has triggered those parts of my creative expression''} both within and outside the performance piece.

This creative expression was exemplified by the inclusion of the performer's paintings as textural elements within the visualisation (shown in Figure \ref{fig:Paintings}).
The performer described the experience of interacting with their own paintings and how it differed from the photographic textures: 
\begin{quote}
\textit{``It was like I was still painting. When I knew my artwork was in the visualisation my artwork expanded, it took me on a continual journey of that artwork. Where with the photos it was completely different... The photos it was more about the subject matter, the content, the city, the nature and transforming that. Where with the artwork it's more about continuing to do that painting, whatever it became.''} 
\end{quote}

\subsection{Iterative and inclusive design process}
Structuring the collaborative development of the piece around the cyclical process of periods of development, followed by opportunities for exploration, evaluation and reflection, ensured the piece progressed in line with the performer's needs and creative goals.
Valuable insights obtained from rehearsal discussions, as well as analysis of gesture and video recordings, served as catalysts for both technical and creative advancements in the development of new prototype interaction and visualisation systems between each rehearsal.
The rehearsal sessions provided crucial moments to check in with the performer, exploring and evaluating each prototype system to assess the alignment with the performer's style of movement and artistic vision. 
This iterative and inclusive design approach played a pivotal role in ensuring that the interaction system responded effectively to the performer’s movement capabilities and that the visualisation system enhanced the performer's existing dance practice in line with the creative imperatives of the piece.

\subsection{Expressing the individuality of movement}
Rather than designing an interaction system for use by anyone, the interaction system created for this work was designed around an individual performer and the unique style of movement style they express within their dance practice. 
This approach of designing an interaction system to express the uniqueness of an individual’s style of movement proved an effective method to create an inclusive interaction system that responded to the movement capabilities of a performer living with physical disability. 
The performer’s condition, cerebral palsy, is a disorder that affects their ability to move and maintain balance and posture.
Specific qualities of the performer's movements were recorded and analysed to ensure the system afforded the performer an expressive interaction experience. 
The range of motion for each sensor placement on the performer's body was recorded, analysed and later mapped within the interaction system to ensure each sensor location could achieve the full range of creative expression. 
Similarly, the pace of movement data expressed at each sensor location was examined to further tailor the interaction experience to the performer’s style of movement. 

The performer explained that their body experiences involuntary movements (spasms), stating \textit{``I never do the same movements because my spasms work in their own way, their own time, their own pace''}. 
The performer described their relationship with their body as an \textit{``improvisation''}, acknowledging and incorporating their involuntary movements as a creative device within their dance practice. 
The decision to randomise the visual texture selection within the piece emerged from these discussions about bodily control and the need to be open to working with unexpected factors that arise in the performer's movement practice, with the performer stating \textit{``I’m not always controlling what I’m doing, I’m just moving''}.
Due to their cerebral palsy, the performer mentioned lacking internal feedback mechanisms to discern the extent of a spasm. 
The presence of the visualisation system provided a mechanism for the performer to visualise involuntary movements, an empowering experience that allowed the performer to better understand their body's movements.
This unforeseen result emerged through the exploration of the creative potential of movement technology, reinforcing the practice-based methodology as a means to explore new forms of corporeal computation.

\subsection{Encouraging new ways of moving through technology}
The decision to place the wearable motion sensors on the performer's lower legs, left wrist and right upper-arm was made with the performer and had a profound impact on the way they made movement decisions within the improvised piece. 
The performer has varying levels of control over different parts of their body, and the choice to wear the sensors at these locations allowed them to explore their body's movement in new and exciting ways.
Wearing the sensors on the performer's lower legs allowed them focus and explore expressive movements with a part of their body that is often overlooked:
\begin{quote}
\textit{``My legs, often they miss out a lot on my movements that are intentional. Because I'm sitting a lot and I move a lot with my upper body. It was really fun to reverse that and focus on what movement I can do with my feet, my toes, my knees, my side, my core. When I lay on my stomach, what do my legs do.''}
\end{quote}
Locating a sensor on the performer's right upper-arm encouraged and challenged them to move a part of their body over which they have limited control:
\begin{quote}
\textit{``It has helped me to recreate the way my brain works to activate my movement. To think about what movement can I get from that part of my body [motions to upper arm]. It’s almost like I wake up more.''}
\end{quote}

The performer explained that their interactions with the wearable devices and the visualisation system have prompted them to reflect on their movements both within and outside the piece, encouraging new movements within their daily life and in rehearsals for other dance performances. 
In addition to encouraging individual movements of the parts of their body where sensors are worn, the performer stated that the placement of the sensors \textit{``really helped me to work with my whole body''} which is something they stated that they don’t often do:
\begin{quote}
\textit{``It might be in my head, but when it’s on my upper arm I think about the movement of my whole body. When it is on my forearm I only think of movement in between here and here [points from upper forearm to lower forearm].''}
\end{quote}

The inclusion of wearable devices within the performer's practice encouraged the performer to move in new ways, exploring movements that were different to other dance performances they have done in the past. 
The performer described this as not just an extension but a \textit{``new direction''} in their movement practice, stating they \textit{``would really like to keep thinking about what it is we are working on here and how that could be part of my future practice''}. 
This further emphasises the role of movement technology to encourage new ways of moving and realising new forms of creative expression. 

\section{Discussion}
The insights gained from this creative exploration can be generalised for application by other creative practitioners aiming to develop expressive movement interaction experiences.
These include the importance of co-design when creating an expressive system for an individual, the creative opportunities afforded by movement as a mode of interaction, the value of recording interactions in an analysable format, and the significance of designing for the interaction experience rather than isolating the technologies involved.

\subsection{Co-design for expressive experiences}
The performer's distinctive movement style and capabilities were included and embraced through the collaborative creation of a personalised interaction system that responds to the specific movement qualities expressed within their dance practice. 
Regularly engaging the performer during the development of the interaction and visualisation systems through an inclusive design process proved to be an effective approach for realising an expressive interaction experience that enhanced the performer’s movement practice through technology.

\subsection{Creative expression of movement data}
When designing creative systems that respond to movement, there is an inescapable requirement to reduce the phenomenological complexity of embodied movement into a subset that can be represented in the digital data.
This concept is explored in depth by Seb Franklin in \cite{alma9939077177201751}, describing that this process of `digitality’ can lead to modes of exclusion and dispossession. 
We explored this concept in this work through the development of a visualisation system that artistically communicates specific facets of the performer’s movements during an improvised dance performance.
During the development of the work, it became apparent that the selection of which aspects of a performer’s movements to creatively convey within the visualisation was constrained by the choice of interface technology.
The use of a IMU-based wearable device constrained the movement data available to the orientation and acceleration of the performer's movements. 
The visualisation system created for this piece was designed to convey the artistic objectives of the piece while working with the limitations of the movement data available.
It emerged through development of this work that the approach to visually conveying the available movement data was a near limitless creative space, constrained only by the imagination of the creative practitioner. 
Defining creative themes (the duality of the city versus nature) and descriptive metaphors for the interaction experience (painting with light) at the outset of the work, helped to narrow this creative space and converge on an interaction experience that achieved the creative imperatives of the work.

\subsection{Recording movement interactions}
The iterative development cycle through which this performance piece emerged was structured around regular evaluation of prototype systems during the rehearsal sessions. 
The performer’s interactions with the prototype systems were observed and discussed, offering valuable insights that would guide and drive future development cycles.
In addition to observation and discussion, multi-modal recordings that simultaneously captured video and gestural data were captured during the rehearsal sessions, and later analysed to reveal insights that would have otherwise been missed in the flow of the session.
The significance of documenting the performer’s interactions in a format that could be later analysed outside of the rehearsal sessions was critical to creating a system that afforded expressive interactions for the performer’s movements.

\subsection{Designing the experience}
Through the iterative development and evaluation of the interaction and visualisation systems, it became evident that while these two systems were distinctive and independent in software, for the performer these two systems were experienced as one.
This highlighted the importance to holistically evaluate the system as it is experienced by the performer.
While many insights can be gained through analysis of the movement data in isolation, the refinement of interaction parameters is most effectively done with the performer present to give direct feedback about their experience of the interaction.
This emphasis on the performer’s interaction experience proved crucial to ensuring that the performer's needs were met and that the experience evolved in line with their creative themes for the piece.

\section{Conclusion and Future Work}
The development and showing of this collaborative performance work has provided a number of insights into the role of technology when working with movement as a mode of interaction.
By augmenting a performer's existing dance practice with an interactive visualisation that responds to their movements, this creative project worked as a `testbed’ for new ideas in movement computing and opened new avenues for the performer's creative expression. 

The research presented in this paper details a in-depth case study into the personalisation of movement interaction systems. 
This work forms one part of a series of case studies, exploring the design of personalised interaction experiences for a diverse range of performers.
The insights described in this paper will be carried forward into future case study collaborations.
The broader research project aims to outline a series of design principles that can be utilised when creating digital instruments to efficiently enable customisation to any individual's movement capabilities and interaction style. 

\bibliographystyle{ACM-Reference-Format}
\bibliography{references}

\appendix

\end{document}